\documentclass[runningheads]{svmult}

\usepackage{makeidx}   % allows index generation
\usepackage{graphicx}  % standard LaTeX graphics tool
\usepackage{subeqnar}  % subnumbers individual equations
\usepackage{multicol}  % used for the two-column index
\usepackage{physprbb}  % modified textarea for proceedings,
\makeindex             % used for the subject index
\begin{document}
\title*{The Masses of Distant Galaxies from Optical Emission Line
Widths}
\toctitle{The Masses of Distant Galaxies from Optical Emission Line Widths}
%\protect\newline in the Particle Deflection Plane}
% allows explicit linebreak for the table of content
%
%
\titlerunning{Optical Emission Line Widths}
% allows abbreviation of title, if the full title is too long
% to fit in the running head
%
\author{Elizabeth Barton Gillespie\inst{1}
\and Liese van Zee\inst{2}}
\authorrunning{Barton Gillespie \& van Zee}
% if there are more than two authors,
% please abbreviate author list for running head
%
%
\institute{Hubble fellow, University of Arizona, Steward Observatory
\\ 933 N. Cherry St., Tucson, AZ, USA 
\and
Indiana University, Department of Astronomy, 727 E. 3rd St., 
Bloomington, IN 47405-7105}
\maketitle              % typesets the title of the contribution

\begin{abstract}
%<= 150 words

Promising methods for studying galaxy evolution rely on optical
emission line width measurements to compare intermediate-redshift
objects to galaxies with equivalent masses at the present epoch.
However, emission lines can be misleading.
We show empirical examples of galaxies with concentrated
central star formation from a survey of galaxies in pairs;
HI observations of these galaxies indicate that the optical line
emission fails to sample their full gravitational potentials.
We use simple models of bulge-forming bursts of star formation
to demonstrate that compact optical morphologies and small 
half-light radii can accompany these anomalously narrow emission lines;
thus late-type bulges forming on rapid (0.5~--~1 Gyr) timescales at 
intermediate redshift would exhibit properties similar 
to those of heavily bursting dwarfs.
We conclude that some of the luminous compact objects observed
at intermediate and high redshift may be starbursts in the
centers of massive galaxies and/or bulges in formation.

\end{abstract}

\section{Introduction}
Optical emission line widths are potentially important diagnostic
tools for measuring the intrinsic gravitational
masses of galaxies within their optical radii.  Because of the sensitivity
requirements for spatially resolved rotation curves,
large surveys of galaxies at intermediate redshift and studies of
galaxies at high redshift use unresolved or ``integrated''
emission line widths, computed from Gaussian fits to emission lines
in the spectrum of the whole galaxy.  However, the results are
sometimes ambiguous in the case of compact star-forming galaxies.

The luminous compact blue galaxies observed at intermediate redshift
(e.g., Koo et al. 1994; 1995)  
have small half-light radii (R$_{\rm e}=1$~--~3.5 kpc) and narrow
emission-line velocity widths ($35 < \sigma < 126$~km~s$^{-1}$).
These properties suggest that although they are
luminous galaxies, compact blue galaxies may be
intrinsically faint galaxies undergoing a strong burst of star formation
(Guzm\'{a}n et al. 1996; 1997).  However, Kobulnicky \& Zaritsky (1999) 
measure high metallicities for
these objects, appropriate only for massive galaxies,
and HST images show possible evidence for 
surrounding older populations (Guzm\'{a}n et al. 1998).
Similarly, at higher redshifts ($z \sim 3$) the ``Lyman break'' galaxies 
also exhibit
narrow integrated line widths that do not correlate with galaxy
luminosity (Pettini et al. 2001).
These observations taken together raise the question of whether 
emission line widths of compact objects accurately trace their
potential wells (Kobulnicky \& Zaritsky 1999).

We present evidence from observations of local galaxies and
simple models of compact star formation that centrally concentrated
star formation changes the measured emission line widths and half-light
radii of galaxies (see also Kobulnicky \& Gebhardt 2000; Pisano et al. 2001).  
This star formation can arise from 
major mergers (Mihos \& Hernquist 1996), 
minor mergers (Mihos \& Hernquist 1994), 
and secular evolution (Pfenniger \& Norman 1990), 
processes that may be directly linked to bulge formation.  The number
of compact blue objects at intermediate redshift that are actually
concentrations of star formation in larger galaxies remains unknown.
If the luminous, compact blue galaxies are frequently bulges in
formation, their number counts contain information about the
timescales for evolution
along the Hubble sequence.

\section{Local Galaxies with Compact Central Star Formation}

In a recent spectroscopic study of the centers of 502 galaxies in 
pairs, Barton, Geller, \& Kenyon (2000) find evidence for correlations
between the star-forming properties of interacting galaxies and the 
pair separations on the sky and in recession velocity.
Their observations are broadly consistent with the 
Mihos \& Hernquist (1996) picture of close galaxy-galaxy
passes triggering gas infall and subsequent star formation in the
central regions of some galaxies.
Barton et al. (2001) examine the Tully-Fisher properties of a subset
of the paired galaxies and find four galaxies that are apparently
overluminous outliers to the relation.
Barton \& van Zee (2001) present VLA HI observations of two of
the outliers; the  radio line widths of the galaxies are substantially broader than 
their resolved optical emission line rotation curves.  Thus, the observations
support the possibility that centrally-concentrated star formation can
give rise to anomalously narrow emission line widths that do not
reflect the full gravitational potentials of the galaxies.

\begin{figure}[h]
\begin{center}
\includegraphics[width=.55\textwidth]{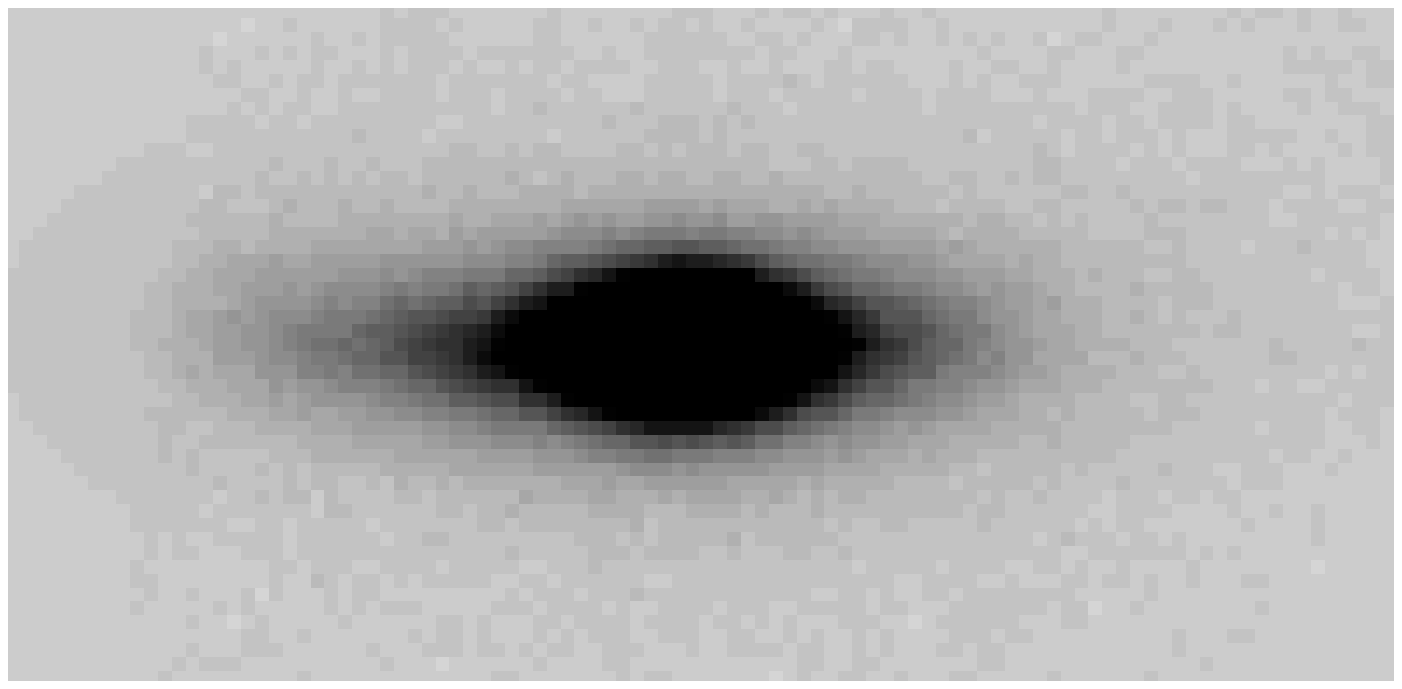}
\includegraphics[width=.55\textwidth]{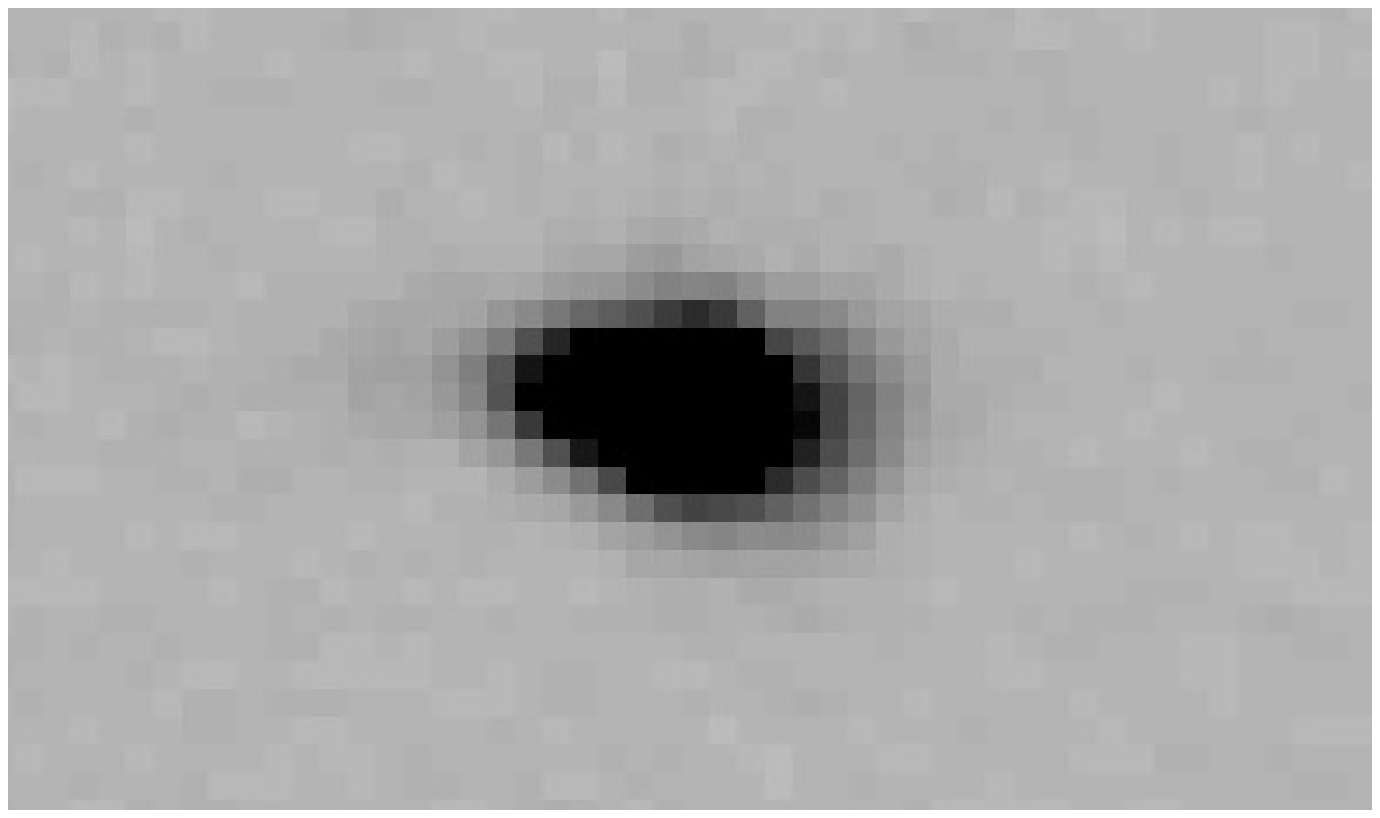}
\end{center}
\caption[]{CGCG 132-062, an interacting galaxy with centrally-concentrated 
optical emission line flux.  We show {\it top} a $B$-band image and
{\it bottom} a longslit spectrum on the same scale, where the horizontal
direction is the spatial direction and the vertical direction is wavelength.  
The emission is largely
confined to the central regions of the galaxy. }
\label{fig:eps1}
\end{figure}

Fig.~\ref{fig:eps1} shows one of the Barton et al. (2001) outliers, 
CGCG 132-062.  The top panel shows the morphology of CGCG 132-062,
including the disk surrounding the central, luminous region.  
The bottom panel shows the
major-axis longslit spectra on the same scale.  The emission is
not spatially extended; it is confined to the central region of
the galaxy and does not include the disky outskirts.

\begin{figure}[h]
\begin{center}
\includegraphics[width=.6\textwidth]{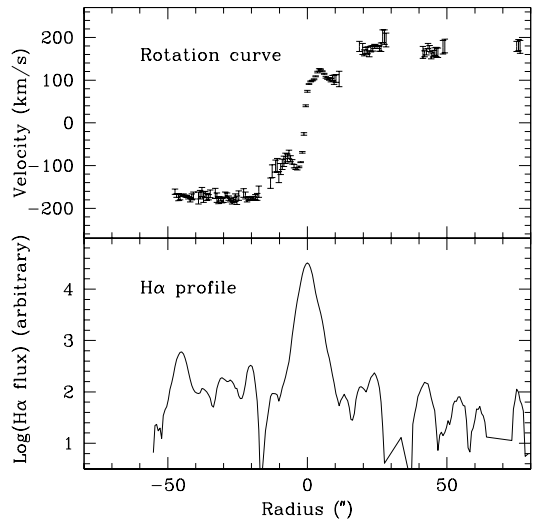}
\end{center}
\caption[]{High surface brightness star formation in the center of
a paired galaxy.  We plot the rotation curve of NGC 470  {\it top}
and the profile of the H$\alpha$ emission incident on the slit {\it bottom}.
The center is $\sim$$100 \times$ higher surface brightness
than the disk but reflects only $\sim$half
of the velocity width of the galaxy.  Thus, although an accurate measurement
at low redshift is possible, neither 
a high redshift spectrum nor an integrated spectrum would reflect the
full velocity width of the galaxy.
}
\label{fig:eps2}
\end{figure}

At higher redshifts, cosmological surface brightness dimming may
render low surface brightness emission from the 
outskirts of a disk invisible.
Thus, for distant galaxies, the observational bias against
measuring full kinematic line widths likely extends to 
galaxies other than the 4 outliers in the Barton et al. (2001) study.
Fig.~\ref{fig:eps2} shows the H-alpha emission profile of
a non-outlier spiral, NGC 470, from Barton et al. (2001).  The
H-alpha in the center of the galaxy is brighter than the outskirts
by a factor of $\sim$100; this central part reflects only
$\sim$50\% of the kinematic width of the galaxy.  An integrated
line profile and perhaps even a 2-dimensional
resolved rotation curve would miss this flux and therefore 
result in an anomalously small line width.

\section{The Effects of Centrally-Concentrated Star Formation}

\begin{figure}[h]
\begin{center}
\includegraphics[width=.6\textwidth]{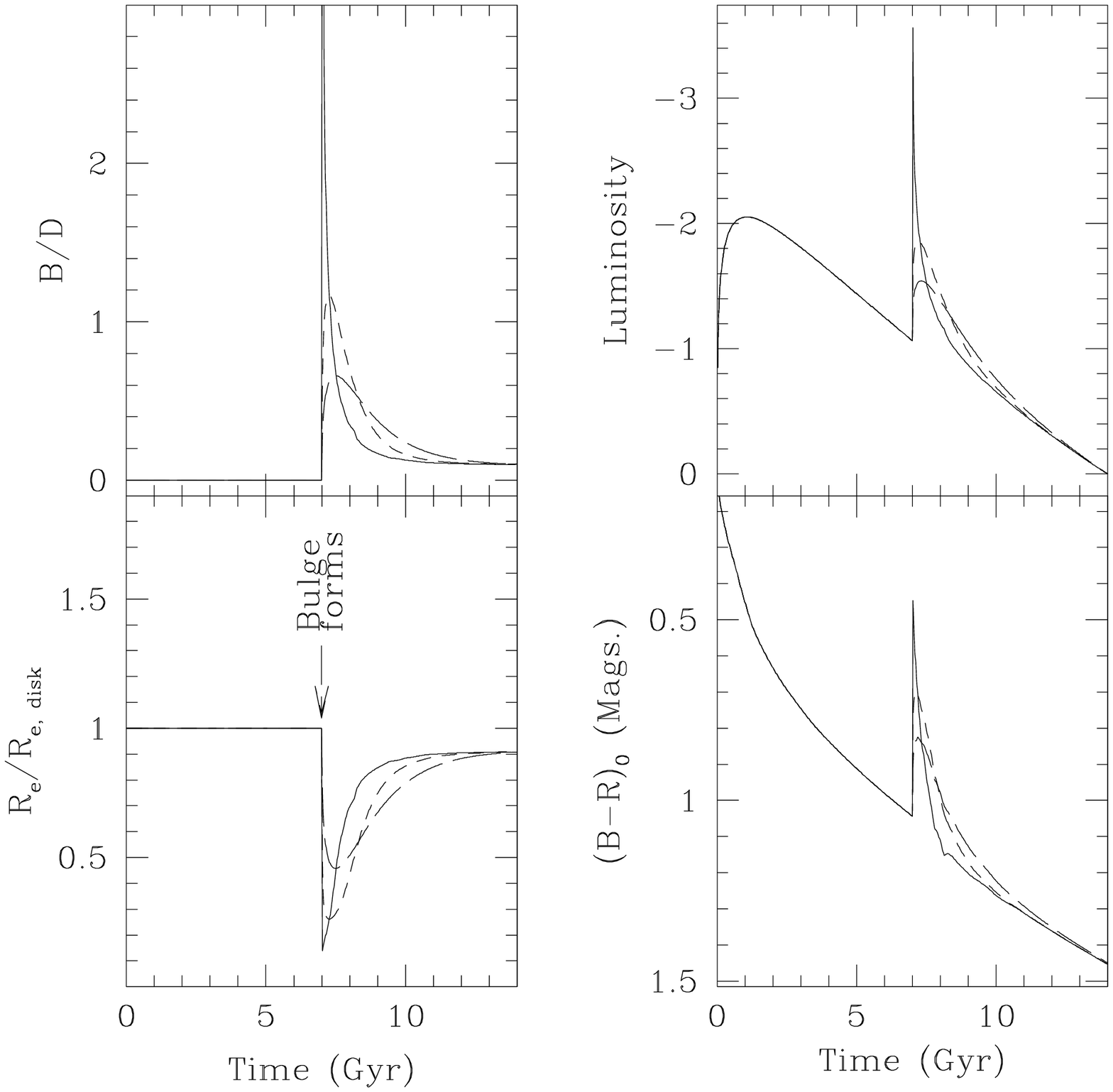}
\end{center}
\caption[]{A simple model for a bulge-forming burst of star formation.  We
plot the bulge-to-disk ratio ({\it upper left}), normalized half-light 
radius ({\it lower left}), normalized luminosity ({\it upper right}), and
total $B-R$ color ({\it lower right}) for a model
spiral galaxy that forms a bulge at intermediate redshift
($T = 7$ Gyr). See the text for more details.}
\label{fig:eps3}
\end{figure}

Large surveys frequently explore the intrinsic 
properties of galaxies using a limited set of
structural parameters 
(e.g., half-light radius, R$_{\rm e}$, or dispersion of
fit to emission lines in the integrated spectrum, $\sigma$).  
However, if star formation rates vary in
different components of the galaxies, the structural parameters of
evolving galaxies will be affected by the differing mass-to-light
ratios in the different components. 
In Fig.~\ref{fig:eps3} 
we expand on the simple model of Barton \& van Zee (2001) to
show that a bulge-forming burst of star formation could
profoundly effect the structural parameters of a galaxy during formation.
Barton \& van Zee (2001) describe the model in detail.
We use the spectral synthesis models of Bruzual \& Charlot
(2001, in preparation) and typical ``exponential bulge'' parameters from
Carollo (1999) and Courteau, de Jong, \& Broeils (1996) 
[final $B/D = 0.1$; radius of disk is $12.5 \times$ radius 
of bulge].  After 7 Gyr, the previously
bulge-less model spiral forms a bulge {\it in situ} in a brief
period of time
(instantaneous: {\it solid line}; 
$\tau = 0.5$~Gyr: {\it short-dashed line}; $\tau = 1$~Gyr:
{\it long-dashed line}).  

The top panels show that, depending on the formation timescale, 
the burst of star formation affects 
many of the basic structural parameters of the galaxy.
During bulge formation, the total luminosity of the $\tau = 0.5$~ or 
1 Gyr models increases by $< 1$ magnitude, but the bulge-to-disk ratio
can peak above 1 and the half-light ratio can dip to 
below 30\% of its original value.
Barton \& van Zee (2001) use same model to show that $\sigma$
for a maximal-disk rotation curve decreases
to as little as 60\% of its original value during bulge formation.

Although the model does not cover every possible evolutionary scenario,
the results are generic and require only formation  within
a relatively short timescale --- less than 1 Gyr --- short enough to allow close galaxy-galaxy
passes, minor mergers, and possibly single-episode secular evolution.
With only $\sim$1~--~2 magnitudes or less of
(transient) luminosity evolution, the temporary
movement in $\sigma$~--~R$_{\rm e}$~ space
is enough to misjudge the nature of these galaxies during bulge formation
(c.f., Fig.~2 in Barton \& van Zee).  Thus, some of the
luminous compact blue galaxies observed at intermediate redshift may
be intrinsically massive galaxies containing compact bursts of star formation.

\section{Luminous Compact Blue Galaxies and Galaxy Evolution}

The luminous compact blue objects, whether dwarfs or spirals,
are clearly some of the most rapidly
evolving galaxies observed at intermediate redshift.
Although extreme examples of these objects are relatively
rare locally, close galaxy-galaxy passes, minor mergers, and secular evolution
may funnel gas into the centers of galaxies more efficiently at higher redshifts, 
where more gas is available.  

An understanding of both dwarf and luminous 
galaxy evolution at intermediate redshift requires
measures of their intrinsic masses and sizes, hence their $z=0$
morphologies and luminosities.  
``Exponential'' bulges have distinctly different properties
from R$^{1/4}$-law bulges (e.g., Andredakis \& Sanders 1994); 
they are candidates for bulges formed via 
secular evolution (Pfenniger \& Norman 1990) or any process that 
sends disk gas into the center of a galaxy.  
If the majority of the luminous, compact blue galaxies are actually
spirals undergoing strong central bursts of star formation, they 
may be consistent with forming exponential bulges.  Thus, 
the fraction of these objects that are actually
bulges in formation may directly reveal the ``exponential'' bulge formation
history of the Universe.

\begin{figure}[h]
\begin{center}
\includegraphics[width=.6\textwidth]{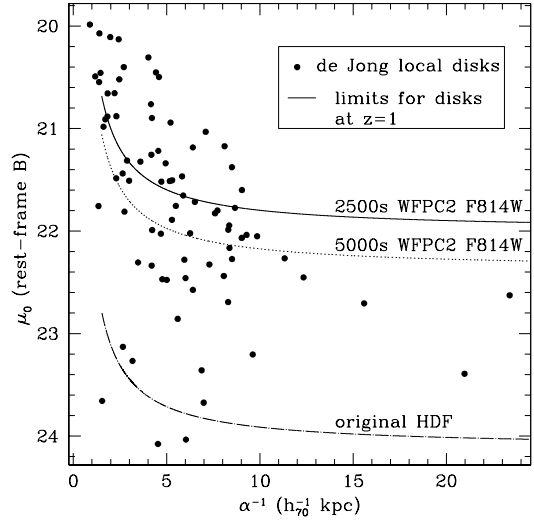}
\end{center}
\caption[]{Detection of face-on disks at $z=1$.  We plot the parameters of
the de Jong (1996) spirals; solid lines show $z=1$ limits with different WFPC2
exposures for accurate measurement
of the parameters of face-on disks under luminous bulges.}
\label{fig:eps4}
\end{figure}

Although existing surveys are not deep enough, 
deep imaging holds promise for distinguishing star 
formation in intrinsically low- and high-mass galaxies.
Fig.~\ref{fig:eps4} shows results from simulations of face-on disks at 
$z=1$ in galaxies with luminous bulges.  The solid lines mark the 
approximate limits at which accurate 
structural parameters measurements are possible.  
The points are local disks from
the de Jong (1996) sample. One to two orbits with WFPC2 are only
enough to characterize the higher surface brightness disks.  Although the
Hubble deep fields detect the majority of even face-on disks
at $z=1$, their combined area is small,
containing the progenitors of few 
late-type galaxies at $z \leq 1$.
Only upcoming ACS surveys will probe deep 
enough over enough area on the sky to detect
the majority of disks that may surround 
luminous, compact blue objects at $z=1$.

%INDEX%%%%%%%%%%%%%%%%%%%%%%%%%%%%%%%%%%%%%%%%%%%%%%%%%%%%%%%%%%%%%%%
% Please check with the editor of your book whether he plans to
% include a "mutual" subject index - if so, please code your entries
% in the standard syntax. For your own purposes you may print your
% "personal" index by using the following commands:
%
%\clearpage
%\addcontentsline{toc}{section}{Index}
%\flushbottom
%\printindex
%%%%%%%%%%%%%%%%%%%%%%%%%%%%%%%%%%%%%%%%%%%%%%%%%%%%%%%%%%%%%%%%%%%%%

\end{document}